\documentclass[twocolumn,aps,amsmath,showpacs,amssymb,nofootinbib,pre]{revtex4}
\usepackage[dvips]{graphicx}
\usepackage[dvips,usenames]{color}
\usepackage{amsmath}

\begin{document}

\title{Robustness of Griffiths effects in homeostatic connectome models}

\author{G\'eza \'Odor}
\affiliation{Research Institute for Technical Physics and Materials Science,
Centre for Energy Research of the Hungarian Academy of Sciences, 
P. O. Box 49, H-1525 Budapest, Hungary}
\date{\today}

\begin{abstract}

I provide numerical evidence for the robustness of the Griffiths phase (GP)
reported previously in dynamical threshold model simulations on a large
human brain network with $N=836733$ connected nodes. The model, with
equalized network sensitivity, is extended in two ways: introduction
of refractory states or by randomized time dependent thresholds.
The non-universal power-law dynamics in an extended control parameter 
region survives these modifications for a short refractory state and
weak disorder. In case of temporal disorder the GP shrinks and for
stronger heterogeneity disappears, leaving behind a mean-field type of
critical transition. Activity avalanche size distributions below the critical 
point decay faster than in the original model, but the addition of 
inhibitory interactions sets it back to the range of experimental values.  

\end{abstract}

\maketitle

\section{Introduction}

Recent experimental evidence suggests \cite{BP03,Chi10}, that our 
brain operates near criticality, where spatial and temporal correlations 
diverge \cite{Hai}, a  dynamics, that enhances the brain 
information-processing capabilities, increasing sensitivity and 
optimizing the dynamic range \cite{Larr}.  
It might be the case that the brain tunes itself near criticality 
via self-organization (SOC) \cite{Bak}, or alternatively, could benefit 
from the same properties if sufficient heterogenities \cite{Lake} 
(that is disorder) are present to induce an extended semi-critical region, 
known as Griffiths Phase (GP) \cite{Griffiths}. 
Alternatively, it has recently been proposed that living systems may 
also self-tune to criticality as the consequence of evolution and
adaptation \cite{adap}. The theory of phase transitions of statistical 
physics applied for these biological models can provide a clue to
better understanding \cite{MArep}.

While SOC models \cite{pruessner} are usually {\it homogeneous}, real 
systems are highly {\it inhomogeneous} and one must consider if the 
heterogeneity is weak enough to apply "clean models" for the description 
of experiments.
Strong disorder can lead to so called rare-region (RR) effects \cite{Vojta}
that smear the phase transitions \cite{round,round2}. 
RR-s are very slowly relaxing domains, remaining in the opposite phase than 
the whole system for a long time, causing slow evolution of the order parameter.
In the entire GP, which is an extended control parameter 
region around the critical point, fluctuations diverge and auto-correlations 
exhibit fat tailed, power-law (PL) behavior, causing burstyness 
\cite{burstcikk}. 
Therefore, here one can observe dynamical critical behavior
without the need of fine tuning. This was proposed to be the reason 
for the working memory in the brain \cite{Johnson} and considered
to be a possible explanation for the observed criticality in 
neuroscience \cite{Hilg-Hutt}.

The seminal experiments by Beggs and Plenz \cite{BP03} reported 
neuronal avalanches 
with size ($s$) dependence, defined as either the number of electrodes with 
supra-threshold activity or the sum of the potentials, according to a power 
law, $p(s) \propto s^{-1.5}$. For the duration distribution of such events
$P(t) \propto t^{-2}$ PL tails were observed. These exponents are in agreement
with the mean-field (MF) exponents of the Directed Percolation (DP) criticality:
$\tau = 3/2$, $\tau_t = 2$, as well as with many other universality classes
sharing the same MF avalanche exponents (see \cite{rmp,MAval}).
Mean-field exponents are expected to occur if the fluctuation effects are
weak, when the system dimensionality is above the upper critical dimension 
$D_c$. For DP $D_c=4$, but recent graph dimension measurements of large
human connectomes suggest lower values \cite{CCcikk}.
Other experimental setups have revealed somewhat different exponents 
\cite{Fried,Shew,Yag} suggesting different universal behaviors.
Furthermore, Palva et al \cite{brainexp} have found that 
source-reconstructed M/EEG data exhibit robust power-law long-range 
time correlations as well as scale-invariant avalanches with a 
broad range of exponents:
$1 < \tau < 1.6$ and $1.5 < \tau_t < 2.4$. 
An obvious explanation for this wide spread of critical exponents 
can be heterogenities, which in the GP cause non-universal dynamical
exponents \cite{Griffiths,Vojta,comment}.

It  has been conjectured that network heterogeneity can cause GP-s 
if the topological (graph) dimension $D$, defined by
$N_r \sim r^D$ ,
where $N_r$ is the number of ($j$) nodes within topological
distance $r=d(i,j)$ from an arbitrary origin ($i$),
is finite \cite{Ma2010}. This hypothesis has been supported 
numerically for various spreading models on scale-free graphs 
\cite{BAGPcikk,wbacikk,basiscikk}. In case of modular topological
structure RR-s are enhanced and Griffiths effects have been reported 
in synthetic brain networks with finite $D$ \cite{MM,Frus,HMNcikk} 
as well as in scale-free networks \cite{Ferr1cikk,MSIScikk}.
However, extended simulations of activity avalanches
on very large human connectomes, possessing 
broad link weight distributions, did not support critical
like phase transition \cite{CCdyncikk}. It turned out that in those
graphs the weight heterogenities are too strong to allow the occurrence of
RR-s. This means that only the strongly connected hubs play a role in 
the activation/deactivation processes and weak nodes just follow them. 
This seems to be rather unrealistic and uneconomic in a brain of billions 
of neurons, thus some kind of input sensitivity equilibration was 
assumed via variable, node dependent thresholds. This makes the system 
homeostatic and simulations proved the occurrence of criticality,
as well as Griffiths effects. 
Indeed there is some evidence that neurons have a
certain adaptation to their input excitation levels \cite{neuroadap}
and can be modeled by variable thresholds \cite{thres}.
Recent theoretical studies have also suggested that homeostatic 
plasticity mechanisms may play a role in facilitating
criticality \cite{65,66,67,68}.
Very recently comparison of modeling and experiments arrived at
a similar conclusion: equalized network sensitivity improves the
predicting power of a model at criticality in agreement with
the FMRI correlations \cite{Rocha2008}. 

A two-state (active/inactive) stochastic threshold model with 
normalized input sensitivity \cite{CCdyncikk}, was investigated by 
simulations on a large, weighted human CC and the spatio-temporal 
behavior of activity avalanches was determined in the inactive phase.
Numerical evidence has been found for a generic scale invariance in 
an extended control parameter space.  
Since this model may look to be too simplistic, 
especially for neuroscientists, now I extend it and perform numerical 
investigations of its dynamical behavior.
The extensions are done in two directions. The addition of refractory
states to the binary active/inactive model and the possibility of
time dependent thresholds. The refractory states happen for one
time step following the active state of a node, thus an activated
neighbor cannot activate back immediately, altering the spreading
process. This is a common feature of brain models \cite{Kandel}.

Time dependency of the network, which can be the consequence
of some external fields, chemicals or some cellular material 
distributed by the blood flow has been simulated by 
altering the thresholds randomly, in a synchronous way for all nodes.
Such temporal disorder, has been shown to lead to temporal 
Griffiths phases, resulting in algebraic size dependency
(space-time flipped) \cite{Vaz11,Voj15} and enormous density 
fluctuations, characterized by an infinitely broad probability 
distribution at criticality. The combination of uncorrelated 
spatial and temporal disorder is irrelevant for the universal
critical behavior \cite{rmp}, but correlated, diffusive 
spatio-temporal disorder has been found to be a relevant 
perturbation \cite{D09,VD16}. In this case I investigate the
effect of long-range correlated spatio-temporal heterogeneity
in the form of the addition of a time dependent global sensitivity
threshold to the quenched spatial disorder of the network.  

\section{Models and methods}

This study uses data from the Open Connectome project (OCP) \cite{OCP}, 
obtained by Diffusion Tensor Imaging (DTI) \cite{DTI} showing 
{\it structural brain connectivity} of the white matter.
Other research on the structural networks have been much smaller sized.
For example the network obtained by Sporns and collaborators, using
diffusion imaging techniques \cite{29,30}, consists of a highly 
coarse-grained mapping of anatomical connections in the human brain,
comprising $N = 998$ brain areas and the fiber tract densities between them.
The graph used here comprises $N=848848$ nodes, allowing one to run
extensive dynamical simulations on present days CPU/GPU clusters,
still large enough to draw conclusions on the scaling behavior
without the effects of finite sizes. Smaller systems near a critical 
phase transition point, where correlations diverge, suffer from finite 
size corrections, which can hide true scaling region and RR effects.

These DTI connectomes of the human brain possess 1 $mm^3$ resolution, 
using a combination of diffusion weighted, functional and structural
magnetic resonance imaging scans. These graphs are symmetric, weighted 
networks, where the weights measure the number of fiber tracts between nodes.
The large graph "KKI-18" used here is generated by the MIGRAINE method,
described in \cite{MIG}. These OCP graphs exhibit hierarchical levels 
by construction from the Desikan cerebral regions with (at least) 
two quite different scales. 
As it was discussed in \cite{CCdyncikk} such coarse grained CC-s suffer 
possible sources of errors, like unknown noise in the data generation; 
underestimation of long connections; radial accuracy, influencing 
endpoints of the tracts and hierarchical levels of the cortical organization;
or transverse accuracy, determining which cortical area is connected 
to another. Still important modifications, such as inhibitory links, 
directedness, or random loss of connections up to $20\%$ confirmed
the robustness of Griffiths effects, suggesting that fine network details 
may not play an important role. It is also important, that the PL tail in
the weight distribution is similar to what was obtained by a synaptic
learning algorithm in an artificial neural network \cite{scarp18}.

The investigated graph exhibits a single giant component of size 
$N = 836733$ nodes (out of $N = 848848$), $41 523 931$ undirected edges
and several small sub-components, ignored here. In \cite{CCcikk} we
found that contrary to the small world network coefficients OCP
graphs exhibit topological dimension slightly above $D=3$ \cite{CCcikk}
and certain amount of universality, supporting the selection of
KKI-18 as a representative of the large human CC-s available. 
This dimensionality suggests weak long-range connections, in addition to the 
$D=3$ dimensional embedding ones and warrants to see heterogeneity 
effects in dynamical models defined on them.  
Further details are discussed in \cite{CCcikk} and \cite{CCdyncikk}.

In my previous study \cite{CCdyncikk} a two-state ($x_i = 0 \ {\rm or} \ 1$) 
spreading model was used to describe the propagation, branching and 
annihilation of activity on this network. 
This model is similar to those of Refs.~\cite{KH,Hai}, a discrete
time stochastic cellular automata (SCA).
The dynamical process, mimicking neuronal avalanches, is started by 
activating a randomly selected node.
At each network update every ($i$) node is visited and tested if the
sum of incoming weights ($w_{i,j}$) of active neighbors reaches a given 
threshold value
\begin{equation}
\sum_{j} x_j w_{i,j}  > K \ .
\end{equation}
If this condition is met a node activation is attempted
with probability $\lambda$. Alternatively, an active node is 
deactivated with probability $\nu$.
New states of the nodes are overwritten only after a full network
update, i.e. a synchronous update is performed at discrete time
steps. The updating process continues as long as there are active 
sites or up to a maximum time limit $t = 10^6$ Monte Carlo sweeps (MCs).
In case the system has fallen into inactivity, the actual time step is
stored to calculate the survival probability $p(t)$
of runs. The average activity: $\rho(t) = 1/N \sum_{i=1}^N x_i$ and the number
of activated nodes during the avalanche $s = \sum_{i=1}^N \sum_{t=1}^T x_i$
of duration is calculated at the end of the simulations. 
This SCA type of synchronous updating is not expected to affect the 
dynamical scaling behavior \cite{HMNcikk} and provides a possibility for 
parallel algorithms. In fact the code was implemented on GPU-s,
with $~ 12\times$ speedup with respect to good CPU cores \cite{GTCposter}.
Measurements on $10^6$ to $10^7$ independent runs, started
from randomly selected, active initial sites were averaged over 
for each control parameter value investigated.

In \cite{CCdyncikk} I showed that this model running on the KKI-18 
graph does not exhibit criticality, but due to the large weight 
differences and graph dimensionality hubs cause a catastrophic, 
discontinuous like transition. To keep the local sustained activity 
requirement for the brain \cite{KH} and avoid nodes, that 
practically do not affect anything I modified the model with 
variable thresholds of equalized sensitivity,
by normalizing the incoming weights as
$w'_{i,j} = w_{i,j}/\sum_{j \in {\rm neighb. of} \ i} w_{i,j}$.

Now I extend this model with the addition of a refractory state,
which occurs with nodes between an activation and deactivation
event for one time step. This prevents activated neighbors from
immediate reactivation of the source and one can expect to find
propagating fronts, resembling dynamical percolation \cite{rmp}.

Another modification is the possibility for allowing a stochastic 
time dependence in the threshold value, by lowering $K$ to $K-\Delta K$, 
with probability $1/2$ at each time MC step for all nodes. 
I tested here  $\Delta K = 0.01, 0.02, 0.04, 0.05$ disorder strengths.

By varying the control parameters, $K$, $\lambda$ and $\nu$ one can find 
a critical transition between an active and an absorbing steady state. 
At a critical point the survival probability of the activity avalanches
(measured by the total number of active sites) is expected to scale 
asymptotically as
\begin{equation}\label{Pscal}
p(t) \propto t^{-\delta} \ ,
\end{equation}
where $\delta$ is the survival probability exponent \cite{GrasTor}.
This is related to the avalanche duration scaling
$P(t) \propto t^{-\tau_t} $ , via the relation  \cite{MAval}:
\begin{equation}\label{taut-del}
\tau_t = 1 + \delta \ .
\end{equation}
In case of single active node initial conditions the number of 
active sites (measured by the total number of active sites during
the spreading experiment) initially grows as
\begin{equation}\label{Nscal}
N(t) \propto t^{\eta} \ ,
\end{equation}
with the exponent $\eta$, which can be related to the avalanche size 
distribution 
\begin{equation}
p(s) \propto s^{-\tau},
\end{equation}
via the scaling relation
\begin{equation}\label{tau-del}
\tau=(1+\eta+2\delta)/(1+\eta+\delta) \ ,
\end{equation}
This was derived in \cite{MAval}, assuming a bell shaped distribution
$p(s,t)$, which is the conditional probability of
an avalanche having size $s$, given it dies at time $t$.

\section{Dynamical simulation results}

In \cite{CCdyncikk} I compared results obtained on directed,
randomly diluted edge variants of the KKI-18 network with those of the 
undirected graph and showed robustness of the GP for those changes. 
For simplicity I consider here the original KKI-18 network, which can 
be obtained from the graph, generated by the MIGRANE algorithm, 
after removing components, which do not belong to the largest connected one.
The control parameter space was restricted by fixing 
$\lambda\simeq 1$, which mimics an efficient brain model and
$\nu=1$ or $\nu=0.95$, corresponding to weak stochasticity in
the deactivation transition. To see critical phase transition 
this requires a threshold value $K \simeq 0.2$.
The time dependent simulations were performed by selecting active 
initial sites randomly and by averaging over $10^7$ realizations 
up to $t_{max}=3\times 10^5$ Monte Carlo sweeps (MCs). 
Throughout the paper time is measured in MCs.

\subsection{Refractory state variable threshold model}

\begin{figure}[h]
\includegraphics[height=5.5cm]{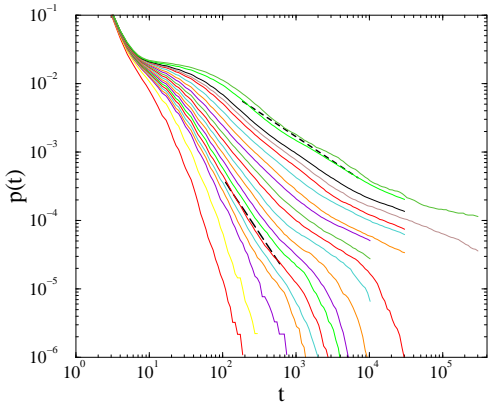}
\caption{\label{s2wlsW} Survival probability of the activity avalanches in 
case of refractory states. Parameters: $K=0.2$, $\nu = 1$ and different 
$\lambda = 0.75$, $0.8$, $0.83$,
$0.84$, $0.85$, $0.86$, $0.87$, $0.88$, $0.89$, $0.9$, $0.91$, $0.92$,
$0.93$, $0.94$, $0.95$, $0.96$, $0.97$, $0.98$, $0.985$ (bottom to top).
The dashed lines show PL fits for the tails of the curve: $\lambda=0.86$ 
as: $\sim t^{-1.6(2)}$ and for the curve: $\lambda=0.98$ as $\sim t^{-0.70(4)}$. }
\end{figure}

Spreading simulations were run at $K = 0.2$ for different $\lambda$ values. 
Fig.~\ref{s2wlsW} suggests a phase transition at $\lambda \simeq 0.985(4)$, 
above which $p(t)$ curves evolve to finite constant values. 
It is very hard to locate the transition clearly, since the evolution slows 
down and log-periodic oscillations also add, as the consequence of the 
modular network. 
Below the transition point we can find $p(t)$ decay curves with truncated
PL tails, characterized by the exponents $0.7 < \delta < 1.6$, as we vary 
$\lambda$ between $\simeq 0.84$ and $0.98$. However, modulation of these PL-s
due to the modular network topology is also clearly visible.
The lower-bound of the scaling region is rather uncertain and estimated 
by the $\lambda$ value, where $p(t)$ and $\rho(t)$ seems to decay faster
than a PL all for all times.
Using the scaling relation (\ref{taut-del}) this means $1.7 < \tau_t < 2.6 $,
falling in the range of brain experiments: $1.5 < \tau_t < 2.4$ \cite{brainexp}. 

\begin{figure}[h]
\includegraphics[height=5.5cm]{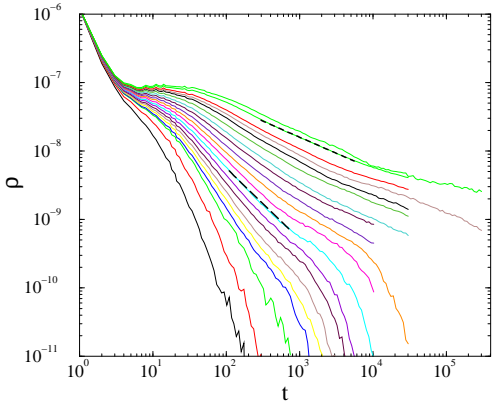}
\caption{\label{rhowlsW} Activity density in case of refractory states.
Parameters: $K=0.2$, $\nu = 1$ and different $\lambda = 0.75$, $0.8$, $0.83$, 
$0.84$, $0.85$, $0.86$, $0.87$, $0.88$, $0.89$, $0.9$, $0.91$, $0.92$, 
$0.93$, $0.94$, $0.95$, $0.96$, $0.97$, $0.98$, $0.985$ (bottom to top). 
The dashed lines show PL fits for the tails of the curve: $\lambda=0.86$ as
$\sim 1/t$ and for the curve: $\lambda=0.98$ as $\sim t^{-0.47(1)}$. }
\end{figure}

One can obtain even clearer PL tails for the density of active sizes $\rho(t)$
with smaller modulations (see the Fig.~\ref{rhowlsW}). 
The GP seems to stretch from $\lambda=0.88$ with a linear deity decay to
$\lambda=0.98$, characterized by a PL fit for the tail $t^{-0.47(1)}$.

\begin{figure}[ht]
\includegraphics[height=5.5cm]{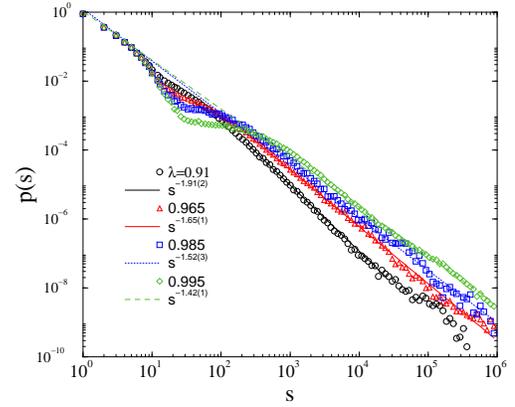}
\caption{\label{elo-t2wlsw} Avalanche size distribution in the
relative threshold model with refractory states, for $K=0.2$, $\nu=1$
and $\lambda= 0.91, 0.965, 0.985, 0.995$ (bottom to top symbols). 
Lines: PL fits for $10^2 < s < 10^5$, for these curves as shown by the legends.}
\end{figure}

Finally, the avalanche size distributions (Fig.~\ref{elo-t2wlsw}), exhibit
clear PL tails, with continuously changing exponents in the range
$1.4 < \tau < 1.91 $, again in agreement with the brain experiments
$1 < \tau < 1.6$ \cite{brainexp}.
The results for the exponent $\tau$ can be crosschecked using the
relation (\ref{tau-del}). For $\lambda=0.98$ one obtains:
$\tau = 1.34(15)$, close to, but a little below, the direct measurements 
of avalanches: $\tau = 1.50(5)$.

\subsection{Time dependent thresholds}\label{sec:tdep}

In the time dependent variable threshold model I fixed $\nu=0.95$, $K = 0.25$
and performed spreading simulations using GPU-s, up to $t_{max}=10^6$
MCs. Before the application of the time dependent thresholds the
GPU accelerated algorithm was tested by the reproduction and extension
of the results of \cite{CCdyncikk} up to $10^7$ MCs, up to two orders of
higher values (see Sect.~\ref{App} and \cite{GTCposter}).
Precise estimates for the critical point have been achieved for 
the model with inhibitory links (see Sect.~\ref{App}).

\begin{figure}[ht]
\includegraphics[height=5.5cm]{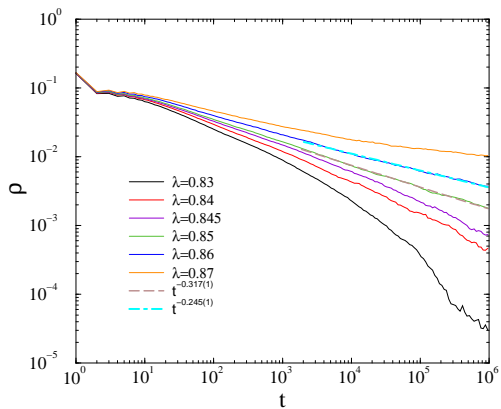}
\caption{\label{swlsW} Average density of activity in the equalized threshold model
with time dependent thresholds started from single active sites. 
Parameters: $K=0.25$, $\Delta K=0.02$, $\nu = 0.95$ and different 
$\lambda = 0.83$, $0.84$, $0.845$, $0.85$, $0.86$, $0.87$ (bottom to top curve). 
Dashed line shows a PL fit $t^{−0.317(1)}$ for the tail of the $\lambda = 0.78$ curve, 
while dot-dashed line shows PL fit $t^{−0.245(1)}$ for the λ$\lambda = 0.86$ curve.
These exponents are away from the values of the three- or four-dimensional
DP criticality, supporting a GP here.}
\end{figure}

In the case of the undirected graph, for $\nu=0.95$, $K = 0.25$ I estimated a 
GP in the region:  $0.85 \le \lambda \le 0.95$ \cite{CCdyncikk}.
Now I investigate what happens if a weak temporal disorder is
added.
Binary distributed variable thresholds have been applied,
first with $\Delta=0.02$ as shown on Fig.~\ref{swlsW}.

The average density decays in a PL manner for $\lambda=0.85$ and  $\lambda=0.86$
with slopes $\alpha = 0.317(1)$,  $\alpha = 0.245(1)$ respectively.
The $\lambda=0.84$ curve seems to be subcritical, while the $\lambda=0.87$ one 
is super-critical.
Lowering the threshold occasionally results in supercritical control 
for $\lambda > 0.86$, which in the time independent case was inside the GP.
But part of the PL region, seems to survive this perturbation.

\begin{figure}[h]
\includegraphics[height=5.5cm]{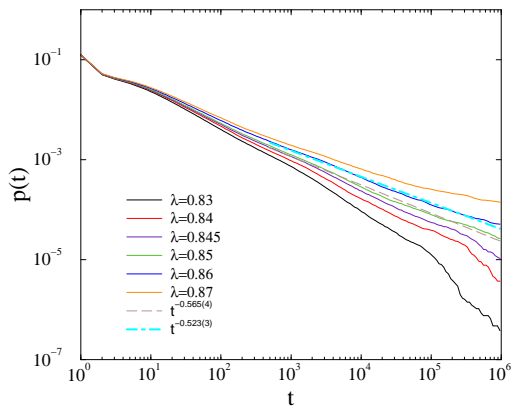}
\caption{\label{s2wlsWi3} Avalanche activity survival probability
in the equalized threshold model with time dependent thresholds started 
from single active sites. Parameters: $K=0.25$, $\Delta K=0.02$, $\nu = 0.95$ 
and different $\lambda = 0.83$, $0.84$,  $0.845$, $0.85$, $0.86$,
$0.87$ (bottom to top curve). Dashed line shows a PL fit $t^{−0.565(4)}$ 
for the tails of the $\lambda = 0.85$, while dot-dashed line shows a PL 
fit $t^{−0.523(3)}$ for the tail of the $\lambda =0.86$ curve.
These exponents are away from the values of the three- or
four-dimensional DP criticality, supporting a GP here.}
\end{figure}

The same conclusion can be drawn by considering the survival probabilities on
Fig.~\ref{s2wlsWi3}, and we can see a narrower GP as before.

\begin{figure}[h]
\includegraphics[height=5.5cm]{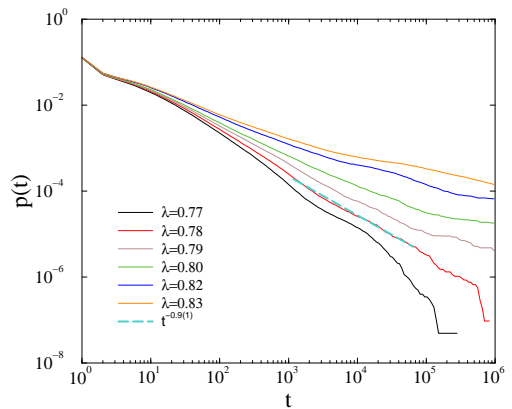}
\caption{\label{Pdep04} Avalanche activity survival probability in the 
equalized threshold model with time dependent thresholds started from 
single active sites. Parameters: $K=0.25$, $\Delta K=0.04$, $\nu = 0.95$ 
and different $\lambda = 0.77$, $0.78$, $0.79$, $0.80$,
$0.87$ (bottom to top curve). Dashed line shows a PL fit for the tail 
of the $\lambda = 0.77$ curve.}
\end{figure}

In contrast, for stronger disorder: $\Delta=0.04$ one can see a critical
transition without the signs of any GP region. At the critical point
$\lambda_c \simeq 0.78(1)$ one can fit decay: $p(t) \propto t^{-0.9(1)}$
close to the MF value of DP (see Fig.~\ref{Pdep04}). The same is true for
the average density of activity.
This suggests that this temporally applied "disorder" is strong enough 
to move all previous curves lying in the GP into the active phase, 
while the previously sub-critical ones remain in the inactive.

\begin{figure}
\includegraphics[height=5.5cm]{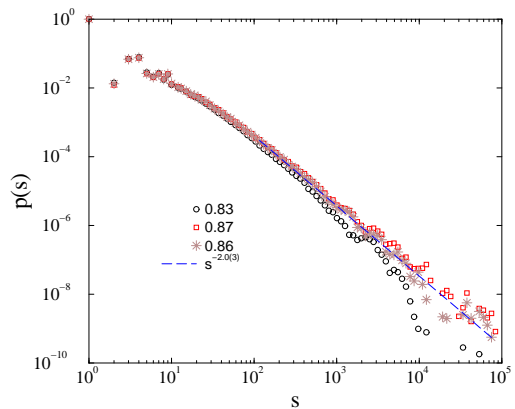}
\caption{\label{elo-t2wlsWi3} Avalanche size distribution in the
relative threshold model with time dependent $K$-s. Parameters:
$K=0.25$, $\Delta K=0.02$ $\nu=0.95$ and $\lambda=0.83, 0.86, 0.87$
(bottom to top symbols). Dashed line: PL fit for $10 < t < 1000$ 
of the  $\lambda=0.86$ (middle) curve.}
\end{figure}

The avalanche size distributions for $\Delta=0.02$ also exhibit 
non-universal PL-s in the GP, as shown in Fig.~\ref{elo-t2wlsWi3},
with somewhat higher slope values: $p(s) \propto s^{-2.0(3)}$.
For clarity I plotted results for just three $\lambda$'s, but
the others fit well to them.

This result for the exponent $\tau$ can be compared with the value we get 
via the scaling relation (\ref{tau-del}). For $\lambda=0.86$ we can obtain: 
$\tau = 1.31(15)$, away from the direct measurement. The reason behind
it must be an asymmetric shaped $p(s,t)$ distribution, which has already 
been showed in \cite{CCdyncikk}. Without this assumption inconsistent 
results were obtained in \cite{inc}.

\subsection{Time dependent thresholds with inhibition}\label{App}

\begin{figure}
\includegraphics[height=5.5cm]{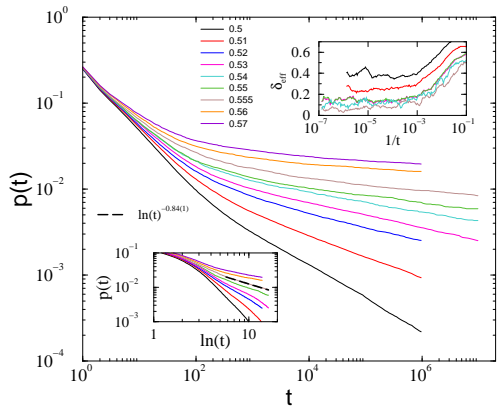}
\caption{\label{ps-0.300.eps} 
Extension of the simulations for the variable threshold model with
inhibitory links of \cite{CCdyncikk}. Avalanche survival decay of the
relative threshold model with $30\%$ inhibitory links at $K=0.1$,
for $\nu = 0.95$ and $\lambda = 0.5$, $0.51$, $0.52$, $0.53$, $0.54$, $0.55$ 
$0.555$, $0.56$, $0.57$, (bottom to top curves).
Lower left inset: The same as the function of $\ln(t)$. Dashed line:
logarithmic scaling fit for the tail of  $\lambda =0.555$ curve. 
Upper right inset: Local slopes of the same curves in opposite order
showing saturation of the nonuniversal exponents in the GP. 
}
\end{figure}

In real brain inhibitory neurons, suppressing communications of
connections exist. I did not simulate such nodes explicitly
but considered negative connections by fast inhibitory inter-neurons.
To model this in \cite{CCdyncikk} I changed $30\%$ of the link 
weights as : $w'_{i,j} = w_{i,j}$ randomly. 
This produces further heterogeneity, thus stronger RR effects. 
Figure~\ref{ps-0.300.eps} shows the survival probabilities,
for $K=0.1$ and $\nu = 0.95$. In this work I show the extension
of the results presented in \cite{CCdyncikk} by two orders of
magnitude in time (from $t_{max}=10^5$ to $t_{max}=10^7$) using
a GPU code, with the primary aim to confirm the parallel program,
and with the secondary aim to provide more precise results at
the critical point and for the GP.

The critical point, above which $p(t)$ signals persistent
activity, is around $\nu=0.56(1)$, hard to locate clearly,
since the evolution slows down and exhibit strong (oscillating)
corrections. Here ultra-slow, log. scaling tail can be observed
of the form: $p(t) \propto \ln(t)^{-0.84(1)}$ as the left inset of
Fig.~\ref{ps-0.300.eps} shows.
This value is smaller than what was found in \cite{CCdyncikk},
for the original, positive weighted graph: $\tilde\delta = 3.5(3)$,
suggesting a lower effective topological dimension due to the
inhibitions. For comparison, in case of spatially quenched disordered 
models belonging to the same universality class one dimension:  
$\tilde\delta = 0.38187$ \cite{Hooy03}, while in two-dimensions: 
$\tilde\delta = 1.9(2)$ \cite{3dsdrg}.
 
Below the transition point the survival exponent changes continuously 
in the range $0 < \delta < 0.5$ as a response for the variation of 
$\nu$ between $0.5$ and $0.57$ (right inset of Fig.~\ref{ps-0.300.eps}).
To see corrections to scaling I determined the local slopes
of the dynamical exponents $\delta$ as the discretized,
logarithmic derivative of (\ref{Pscal}).
The effective exponent of $\delta$ is measured as
\begin{equation}  \label{deff}
\delta_\mathrm{eff}(t) = -\frac {\ln p(t) - \ln p(t') } {\ln(t) - \ln(t')} \ ,
\end{equation}
using $t - t'=8$. A remarkable stabilization of $\delta_{\rm eff}$ can be
observed for $10^4 < t < 10^7$ in the GP.

\begin{figure}
\includegraphics[height=5.5cm]{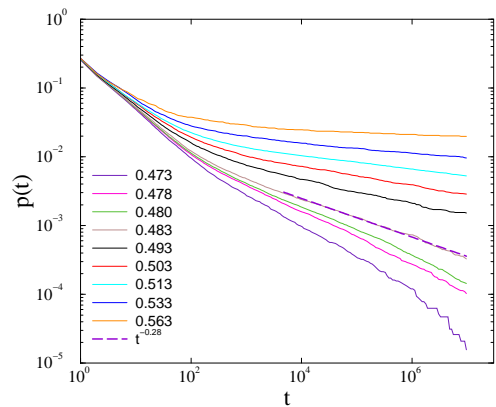}
\caption{\label{ps-0.300-09.eps}
Avalanche survival decay of the relative threshold model with $30\%$ 
inhibitory links at $K=0.1$ and  $\Delta K=0.01$,
for $\nu = 0.95$ and $\lambda = 0.473$, $0.478$, $0.48$, $0.483$, $0.493$, 
$0.503$, $0.513$, $0.533$, $0.563$, (bottom to top curves).
The dashed line shows a PL fit $t^{-0.28(1)}$ for the tail of the λ 
$\lambda = 0.483$ curve.
}
\end{figure}

\begin{figure}
\includegraphics[height=5.5cm]{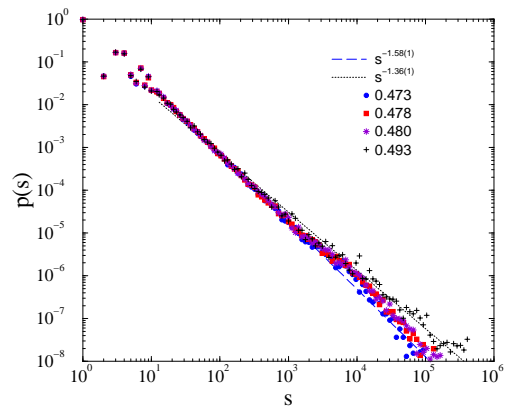}
\caption{\label{elo-pa} Avalanche size distribution of the time dependent
relative threshold model with $30\%$ inhibitory links at $K=0.1$, $\Delta K=0.01$,
$\nu=0.95$ and $\lambda=0.473,0.478,0.480,0.493$ (bottom to top symbols)
Dashed lines: PL fits for the tails of the $\lambda=0.473$ and 
$\lambda=0.493$ curves (bottom to top).}
\end{figure}

As in Section \ref{sec:tdep} I applied time dependent thresholds
by lowering $K=0.1$ to  $K=0.09$ at random time steps with probability
$0.5$. In this case the GP shrunk to the region: $0.483 < \lambda < 0.513$
approximately (see Fig~\ref{ps-0.300-09.eps}). Thus the critical point 
moved down to $\lambda  \simeq 0.51(1)$ with respect to the time independent
model.
Near the critical point the avalanche size distributions exhibit
roughly $p(s) \propto t^{-1.5}$ PL tails in agreement with experiments 
as shown on Fig.~\ref{elo-pa}. Furthermore, the exponents decrease 
slightly by increasing $\lambda$.

Similarly to the non-inhibitory case, for larger temporal disorder,
i. e. for $\Delta K = 0.05$ one cannot see clear PL-s, or maybe just in
a narrow region, hindered by strong periodic modulations. Location of
the phase transition point in this case is hard, but one can estimate it to 
be $\lambda \simeq 0.40(1)$. The avalanche size distributions exhibit 
$p(s) \propto s^{-1.4}$ PL tails here, but extended GP cannot be observed 
now.

\section{Discussion and Conclusions}

I have extended the dynamical scaling study of the activity behavior of
the connectome model started in \cite{CCdyncikk} with the addition of
refractory states and time dependent threshold values. 
High precision simulations showed the survival of GP in both cases,
albeit in case of temporal disorder the GP shrinks. By lowering or 
increasing the threshold occasionally, RR-s of the original GP feel super- or
sub-critical "quenches". Thus their evolutions do not follow the
slow dynamics with large characteristic times.
In the case of the addition of refractory states the GP seems to be even
wider as in the two-state model.
The avalanche size distributions exhibit fat tails with control parameter
dependent PL exponents near the MF value, but the $\tau$ exponent
does not satisfy the scaling law (\ref{tau-del}) connecting it to
$\delta$ and $\eta$ in general, suggesting asymmetric $p(s|t)$
distributions like in \cite{CCdyncikk}.

In summary, while in the case of quasi-static, non-equalized threshold 
models fat tail distributed weights exclude rare-region effects
(thus a GP), in equalized ones, simulations provide numerical 
evidence for robust occurrence of Griffiths effects for 
susceptible-excited-susceptible type of models \cite{CCdyncikk},
for susceptible-excited-refractory-susceptible models, even in the case 
of weak temporal noise. Inhibitory weights do not alter this scenario 
qualitatively, but provide homeostatic states in which the 
scaling behavior is quantitatively more similar to the brain experiments.

In particular, for the susceptible-excited-refractory-susceptible model
the duration exponents in the GP are: $1.7 < τ\tau_t < 2.6$, while the
avalanche size exponents are in the range: $1.4 < \tau τ< 1.91$.
For the weakly time-dependent ($\Delta=0.02$) susceptible-excited-susceptible 
model one can estimate: $1.5 < τ\tau_t < 1.7$ and $2 < \tau τ< 2.28$.
In the case of inhibition, the scaling exponents of the weakly time-dependent, 
susceptible-excited-susceptible model least squares fitting and 
(\ref{tau-del}) results in: $1 < τ\tau_t < 1.5$,  $1.1 < \tau < 1.6$. 
In general, for the durations a minimum value may emerge as: $\tau_t=1+\delta > 1$,
as a consequence of a non-conventional critical point, at the top of the GP 
in case of logarithmic scaling. For the avalanche size distributions there
is no such minimum value of $\tau=2.2$, reported in a fuse model, which is
defined on hierarchical modular networks \cite{fuse}. Instead, the
simulations on this connectome, or in the case of non-hierarchical
modular networks \cite{MSIScikk} $\tau < 1$ can be observed only.
 
Simultaneous, interacting avalanches should not alter the global
universal dynamical critical exponents in statistical physics, although one can
speculate how to define an avalanche in this case. If one could measure
the activity spreading causally, one could decide if activated
nodes are distinguishable or not. In the former case one can apply
a neutral theory and obtain scale-free dynamics even away from the edge of
a phase transition \cite{neutral}. In the latter case I believe one 
should not obtain different critical behavior, like in case of identical 
multi-component reaction-diffusion systems in higher dimensions \cite{rmp}.
In a more realistic model, exhibiting memory of nodes in accumulating
action potentials, interacting avalanches can influence their mutual
spreading behavior, like for a pair-contact process \cite{rmp}
and initial condition dependent scaling may also emerge. 
  
It is important to stress that this study does not aim to exclude
SOC generating synaptic learning and plasticity mechanisms, leading to 
homeostatic states and closeness of a criticality in brain models.
Instead of that I show how heterogeneities alter the dynamical behavior,
causing critical like PL-s in an extended control parameter region
even below the singular phase transition point and provide a possible 
explanation for the spread of exponents measured.
Experimental justification of GP would require to measuring the
avalanche and autocorrelation exponents in high resolution
and show how the scattered exponents depend on some control parameters 
(sensitivity, inhibition, ... etc.) of the brain. 
An interesting continuation of this work would be to test {\it partially} 
equalized input sensitivities, where first order phase transition 
was shown \cite{CCdyncikk}, in agreement with brain
experiments, showing abrupt transition between up and down states
see \cite{updown}.

The codes and the graph used here are available on request from the author.

\section*{Acknowledgments}

I thank Vince Varga for developing the GPU code and appreciate the comments 
from M. Palva, M. Gastner and R. Juh\'asz. The simulations have been performed 
on the NIIF HPC Hungarian national supercomputer machines.
Support from the Hungarian research fund OTKA (K109577 and K128989) 
is acknowledged.


\begin{thebibliography}{50}

\bibitem{BP03} J. Beggs and D. Plenz, {\it Neuronal avalanches in neocortical circuits},
J. Neurosci., {\bf 23}, (2003) 11167.
\bibitem{Chi10} D. R. Chialvo,  Nature Physics, {\it Emergent complex neural dynamics},
{\bf 6}, (2010) 744.
\bibitem{Hai} A. Haimovici, E. Tagliazucchi, P. Balenzuela, and D. R. Chialvo, 
{\it Brain Organization into Resting State Networks
Emerges at Criticality on a Model of the Human Connectome},
Phys. Rev. Lett. {\bf 110}, (2013) 178101.
\bibitem{Larr} D. B. Larremore, W.L. Shew and J. G. Restrepo, 
{\it Predicting Criticality and Dynamic Range in Complex Networks: 
Effects of Topology}, Phys. Rev. Lett. {\bf 106}, (2011) 058101.
\bibitem{Bak} P. Bak, C. Tang and K. Wiesenfeld, 
{\it Self-organized criticality}, Phys. Rev. A {\bf 38}, (1988) 364.
\bibitem{Lake} B. B. Lake et al, 
{\it Neuronal subtypes and diversity revealed by single-nucleus RNA sequencing of the human brain},
Science, {\bf 352}, (2016) 1596.
\bibitem{Griffiths} R. B. Griffiths,
{\it Nonanalytic Behavior Above the Critical Point in a Random Ising Ferromagnet},
Phys. Rev. Lett. {\bf 23}, 17 (1969).
\bibitem{adap} J. Hidalgo et al., {\it Information-based fitness and the emergence
of criticality in living systems}, PNAS {\bf 111}, 10095 (2014).
\bibitem{MArep} Miguel A. M\~noz, {\it Colloquium: Criticality and dynamical scaling in living systems},
Rev. Mod. Phys. {\bf 90} (2018) 031001.  
\bibitem{pruessner} G. Pruessner, {\it Self Organized Criticality},
Cambridge University Press, Cambridge 2012.
\bibitem{Vojta} T. Vojta, {\it Rare region effects at classical, quantum and nonequilibrium phase transitions}, 
J. Physics A: Math. and Gen. {\bf39}, R143 (2006).
\bibitem{comment} Strictly speaking we cannot prove GP in finite models
without the possibility of showing size independence, as in the case of 
the available connectome graphs, but Griffiths effects.
\bibitem{round} P. M. Villa Martin, J. A. Bonachela and M.A. Mu\~noz,
{\it Quenched disorder forbids discontinuous transitions in nonequilibrium
low-dimensional systems}, Phys. Rev. E {\bf 89}, (2014) 012145.
\bibitem{round2} P. M. Villa Martin, M. Moretti and M.A. Mu\~noz,
{\it Rounding of abrupt phase transitions in brain networks}, 
J. Stat. Mech. (2015) P01003.
\bibitem{burstcikk} G. \'Odor, 
{\it Slow, bursty dynamics as a consequence of quenched network topologies}, Phys. Rev. E {\bf 89}, 042102 (2014)
\bibitem{Johnson} S. Johnson, J. J. Torres, and J. Marro, 
{\it Robust Short-Term Memory without Synaptic Learning}, PLoS ONE {\bf 8(1)}: e50276 (2013)
\bibitem{Hilg-Hutt} C. C. Hilgetag and M.-T. Hutt, 
{\it Hierarchical modular brain connectivity is a stretch for criticality.},
Trends in Cog. Sci. {\bf 18} (2013) 114.
\bibitem{MAval} M. A. Mu\~noz, R. Dickman, A. Vespignani and S. Zapperi,
    Phys. Rev. E {\bf 59}, 6175 (1999).
\bibitem{rmp} G.~\'Odor, {\it Universality classes in nonequilibrium lattice systems},
Rev. Mod. Phys. {\bf 76}, 663 (2004).
\bibitem{CCcikk} M. T. Gastner and G. \'Odor,
{\it The topology of large Open Connectome networks for the human brain},
Sci. Rep. {\bf 6}, (2016) 27249.
\bibitem{Fried} N. Friedman, et al, {\it Universal critical dynamics in high resolution neuronal
avalanche data.}, Phys. Rev. Lett. {\bf 108}, 208102 (2012).
\bibitem{Shew} W. L. Shew, W. P. Clawson, J. Pobst, Y. Karimipanah, N. C. Wright, and R. Wessel.
{\it Adaptation to sensory input tunes visual cortex to criticality}, Nat. Phys., {\bf 11}, (2015) 659–663.
\bibitem{Yag} M. Yaghoubi  et al, {\it Neuronal avalanche dynamics indicates different universality
classes in neuronal cultures}, Sci. Rep., {\bf 8}, (2018) 3417.
\bibitem{brainexp} J. M. Palva, A. Zhigalov, J. Hirvonen, O. Korhonen, K. Linkenkaer-Hansen, and S. Palva,
{\it Neuronal long-range temporal correlations and avalanche dynamics are correlated with
behavioral scaling laws}, Proc. Natl. Acad. Sci. USA, {\bf 110}, (2013) 3585–3590.
\bibitem{Ma2010} M.~A. Mu\~noz, R.~Juh\'asz, C.~Castellano and G.~\'Odor,
{\it Griffiths phases on complex networks}, Phys. Rev. Lett. {\bf 105} (2010) 128701.
\bibitem{BAGPcikk}  G.~\'Odor and R.~Pastor-Satorras,
{\it Slow dynamics and rare-region effects in the contact process
on weighted tree networks}, Phys. Rev. E {\bf 86}, (2012) 026117.
\bibitem{wbacikk} G.~\'Odor, {\it Rare regions of the susceptible-infected-susceptible
model on Barab´asi-Albert networks}, Phys. Rev. E {\bf 87}, (2013) 042132.
\bibitem{basiscikk} G.\'Odor, {\it Spectral analysis and slow spreading dynamics
on complex networks}, Phys. Rev. E {\bf 88}, (2013) 032109.
\bibitem{MM} P. Moretti and M. A. Mu\~noz, {\it Griffiths phases and the
stretching of criticality in brain networks},
Nature Communications {\bf 4}, (2013) 2521.
\bibitem{Frus} P. Villegas, P. Moretti and Miguel A. Mu\~noz, Scientific Reports {\bf 4}, 5990 (2014)
\bibitem{HMNcikk} G.\'Odor, R. Dickman and G.\'Odor,
{\it Griffiths phases and localization in hierarchical modular networks},
Sci. Rep. {\bf 5}, (2015) 14451.
\bibitem{Ferr1cikk} W. Cota, S. C. Ferreira and G. \'Odor,
{\it Griffiths effects of the susceptible-infected-susceptible epidemic model on random power-law networks }, 
Phys. Rev. E {\bf 93}, 032322 (2016).
\bibitem{MSIScikk} Wesley Cota, Silvio C. Ferreira, and G. \'Odor, {\it
Griffiths phases in infinite-dimensional, non-hierarchical modular networks},
Scientific Reports {\bf 8}, (2018) 9144.
\bibitem{CCdyncikk} G. \'Odor, {\it Critical dynamics on a large human Open Connectome network},
Phys. Rev. E {\bf 94}, (2016) 062411.
\bibitem{neuroadap} R. Azouz and C. M. Gray, PNAS {\bf 97}, 8110 (2000).
\bibitem{thres} M.-T. H\"utt, M. K. Jain , C. C. Hilgetag, A. Lesne, Chaos, Solitons \& Fractals, {\bf 45}, 611 (2012).
\bibitem{65} Felix Droste, Anne-Ly Do and Thilo Gross, {\it Analytical investigation of 
self-organized criticality in neural networks}, J. R. Soc. Interface 10:20120558 (2013).
\bibitem{66} Gustavo Deco, Adri ́an Ponce-Alvarez, Patric Hagmann, Gian Luca Romani, Dante Mantini, and Maurizio Corbetta. 
{\it How Local Excitation–Inhibition Ratio Impacts the Whole Brain Dynamics}, 
The Journal of Neuroscience, {\bf 34}, (2014) 7886.
\bibitem{67} Peter J. Hellyer, Barbara Jachs, Claudia Clopath, Robert Leech. 
{\it Local inhibitory plasticity tunes macroscopic brain dynamics and allows the emergence of 
functional brain networks}, NeuroImage {\bf 124},(2015) 85.
\bibitem{68} Peter John Hellyer, Claudia Clopath, Angie A. Kehagia, Federico E. Turkheimer, Robert Leech. 
{\it From homeostasis to behavior: Balanced activity in an exploration of embodied dynamic 
environmental-neural interaction}, PLoS Comput Biology 13(8): e1005721 (2017).
\bibitem{Rocha2008} Rodrigo P. Rocha et al, {\it Homeostatic plasticity and emergence of 
functional networks in a whole-brain model at criticality}, Sci. Rep. (2018) 15682.
\bibitem{Kandel} E. R. Kandel, {\it Principles of neural science, 5-th ed.}, 
The MCGraw-Hill Companies, (2013).
\bibitem{Vaz11}  F. Vazquez, J. A. Bonachela, C. L\'opez,  and M. A. Mu\~noz, 
Phys. Rev. Lett. {\bf 106}, (2011) 235702.
\bibitem{Voj15} T. Vojta and J. A. Hoyos, Europhysics Letters {\bf 112}, (2015) 30002.
\bibitem{D09}  R. Dickman, 
{\it A contact process with mobile disorder}, J. Stat. Mech. (2009) P08016.
\bibitem{VD16} T. Vojta, R. Dickman, Phys. Rev. E {\bf 93}, (2016) 032143.
\bibitem{OCP} See: https://neurodata.io
\bibitem{DTI} Bennett A. Landman et al, NeuroImage {\bf 54} (2011) 2854–2866.
\bibitem{29} P. Hagmann, et al. {\it Mapping the Structural Core of
Human Cerebral Cortex.} PLoS Biol. {\bf 6}, e159 (2008).
\bibitem{30} C. J. Honey, et al. {\it Predicting human resting-state
functional connectivity from structural connectivity},
Proc. Natl. Acad. Sci. {\bf 106}, 2035–2040 (2009).
\bibitem{MIG} W.~G.~Roncal et al., {\it MIGRAINE: MRI Graph Reliability Analysis and Inference for Connectomics}, 
in Proceedings of the IEEE Global Conference on Signal and Information Processing 
(John Hopkins University, Maryland, prepint: arXiv.org:1312.4875,
published in: 2013 IEEE Global Conference on Signal and Information Processing, 10.1109/GlobalSIP.2013.6736878.
\bibitem{scarp18} S. Scarpetta, I. Apicella, L. Minati and A. de Candia, Phys. Rev. E {\bf 97}, 062305 (2018).
\bibitem{GTCposter} G. \'Odor and V. Varga, {\it Simulation of whole brain models on large human 
connectomes}, poster preseted in GTC2018 conference San Jose, USA, 2018, 
https://www.nvidia.com/en-us/gtc/poster-gallery/hpc-and-supercomputing.
\bibitem{KH} M. Kaiser and C. C. Hilgetag, {\it Optimal hierarchical modular 
topologies for producing limited sustained activation of neural networks}, Front. in Neuroinf., {\bf 4} (2010) 8.
\bibitem{GrasTor} P. Grassberger and A. de la Torre,
Ann. Phys. {\bf 122}, 373 (1979).
\bibitem{Hooy03} J. Hooyberghs, F. Igl\'oi, C. Vanderzande, Phys. Rev. Lett. {\bf 69}, (2003), 100601.
\bibitem{3dsdrg} I. A. Kov\'acs, F. Igl\'oi,  Phys. Rev. B {\bf 83}, 174207 (2011).
\bibitem{inc} A. Chessa, H. E. Stanley, A. Vespignani, and S. Zapperi, Phys. Rev. E {\bf 59}, (1999) R12.
\bibitem{fuse} Paolo Moretti, Bastien Dietemann, Michael Zaiser, Sci. Rep. {\bf 8}, (2018) 12090.
\bibitem{neutral} M. Martinello, J. Hidalgo, A. Maritan, S. di Santo, D. Plenz, and M. A. Mu\~noz,  
Phys. Rev. X {\bf 7}, (2017) 041071.
\bibitem{updown} D. A. McCormick, {\it Neuronal networks: Flip-Flops in the brain}, 
Curr Biol {\bf 15}, (2005) R294–R296.
\end{thebibliography}
\end{document}